# On-surface synthesis of nitrogen-doped nanographenes with 5-7 membered rings

Dmitry Skidin,[a,b] Frank Eisenhut,[a,b] Marcus Richter,[b,c] Seddigheh Nikipar,[a] Justus Krüger,[a,b] Dmitry A. Ryndyk,[d] Reinhard Berger,[b,c] Gianaurelio Cuniberti,[a,b,e] Xinliang Feng,[b,c] Francesca Moresco*[b]

**We report on the formation of nitrogen-doped nanographenes containing five- and seven-membered rings by thermally induced cyclodehydrogenation on the Au(111) surface. Using scanning tunneling microscopy and supported by calculations, we investigated the structure of precursor and targets, as well as of intermediates. Scanning tunneling spectroscopy shows that the electronic properties of the target nanographenes are strongly influenced by the additional formation of non-hexagonal rings.**

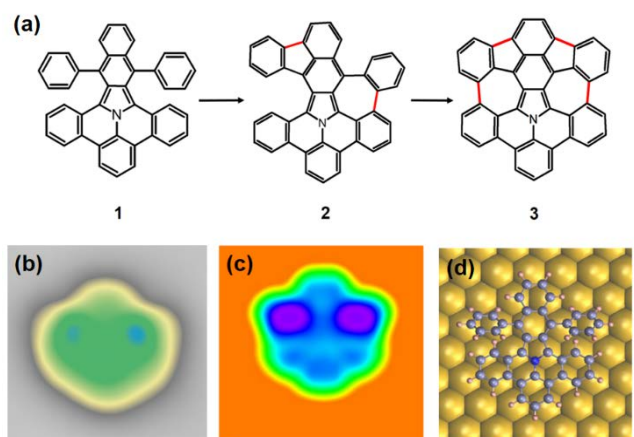

Structural defects are known to significantly affect the electronic and chemical properties of graphene, making it possible to achieve new functionalities.[1] Microscopy studies reveal that graphene can contain rings with an anomalous number of carbon atoms. Pentagon-heptagon defects, so called azulene motifs, for example, strongly perturb the sp²-carbon framework, and the band dispersion at the Fermi level is drastically modified leading to an increased density of states close to the Fermi-level.[2-4] Furthermore, the introduction of heteroatoms like nitrogen provides an efficient way to tune the chemical and electronic properties of graphene.[5-7]

Extended polycyclic aromatic hydrocarbons (PAHs), also namely nanographenes, can be regarded as atomically precise cutouts of graphene.[8] They can serve as model system for understanding the role of single structural defects and heteroatoms in a graphene-like structure.

Recently, the introduction of different defects has been reported for carbon nanotubes[3,9] and several groups obtained nanographenes with non-hexagonal rings embedded.[10-13] Moreover, heteroatom-doped graphene nanoribbons were successfully synthesized using nitrogen-doped precursors,[14] demonstrating p-n junctions[15] and the tuning of the band gap.[16] Molecular nanostructures and nanographenes with tailored properties can be obtained by chemical reactions performed directly on-surface.[17,18] In particular, surface-assisted cyclodehydrogenation reactions lead to the synthesis of unprecedented nanographenes. Although the formation of hexagonal rings has already been successfully achieved for several systems, it has been recently shown that the integration of rings with unusual number of carbon atoms (*i.e.* 4, 5, 7, 8) can be obtained by on-surface reactions of carefully designed precursors.[19-23] Notably, the implementation of azulene motifs containing 5 and 7-membered rings was demonstrated in nanographenes after the cyclodehydrogenation reaction.[24-26]

Here, we describe the on-surface chemistry approach used to create atomically-defined 5-7 defects in nitrogen-doped nanographenes. To the best of our knowledge, this is the first example of a N-doped nanographene containing heteroatom-free 7-membered rings on surface. Starting from a benzoisoindole precursor (**1** in Fig. 1a), we employed thermally-assisted cyclodehydrogenation reactions to obtain the fully dehydrogenated nitrogen-doped nanographene **3**. Possible different intermediate species can be formed along the reaction path (one containing one 5-membered and one 7-membered ring can be seen as intermediate **2**). We used scanning tunneling microscopy (STM) and spectroscopy (STS) at low temperature to resolve the structure of the different molecular species adsorbed on the Au(111) surface, and study the electronic resonances of the precursor **1** and the final nanographene **3**.

Fig. 1 Precursor molecule **1** after deposition on the Au(111) surface. (a) Proposed on-surface reaction path. The initial molecule **1** undergoes thermally-assisted cyclodehydrogenation to form the target **3** with 5-7 defects through one possible intermediate step **2**. (b) STM image of a single precursor molecule **1** (U = 0.5 V, I = 100 pA, 2.5 nm x 2.5 nm). (c) Calculated STM image of the precursor **1** (the same size). (d) Atomistic model of the precursor **1** adsorbed on the Au(111) surface after geometry optimization using DFT. Color code: gray – carbon, pink – hydrogen, blue – nitrogen.

We commenced our study with the synthesis of the precursor 5,17-diphenylbenzo-[7,8]naphtho-[2',3':1,2]indolizino-[6,5,4,3-*def*]phenanthridine **1** starting from 8*H*-isoquinolino[4,3,2-*de*]-phenanthridin-9-ium tetrafluoroborate based on a published protocol (**S1**, details in the supplementary information).[27] After depositing the precursor **1** on an atomically clean Au(111) surface held at room temperature at a sub-monolayer coverage, the molecule **1** mostly adsorb at the elbow sites of the Au(111) surface reconstruction. In the STM images (Fig. 1b) the precursor molecules **1** show a characteristic shape with two bright features on both sides of the symmetry axis. These features correspond to the substituted phenyl rings, and indicate that the molecules **1** are not adsorbed flat on the surface. On a supramolecular scale, such non-planarity results in the formation of chain-like tetrameric structures, stabilized by non-covalent interactions (see S10 in the supplementary information).[28]

The adsorption of individual molecules and the molecular geometry on the Au(111) surface were modelled using density functional theory (DFT)[29,30] as implemented in the CP2K software package with a Quickstep module.[31] We applied Perdew-Burke-Ernzerhof (PBE)[32] exchange-correlation functional, with Goedecker-Teter-Hutter (GTH)[33] pseudo-potentials and a valence double-ζ basis set. Dispersion corrections were included through the standard D2 Grimme parameterization.[34] The geometry optimization for every intermediate state (initial, partial dehydrogenated and final structure) was carried out with the convergence criteria of $10^{-5}$ Hartree for the SCF energy, while the gold substrate was modeled by three layers with the lattice constant of 2.88 Å and the first layer was fixed and the two top layers including molecules were relaxed. To calculate the STM topography images, the Tersoff-Hamann method was employed.

In Fig. 1c, the calculated STM image of the adsorbed precursor molecule **1** is shown. It is in good agreement with the experimental one and reproduces the molecular shape with the higher appearing phenyl rings on the two sides. The twisting of the phenyl rings upon adsorption of the nanographenes **1** on Au(111) is confirmed by the DFT optimization (Fig. 1d).

To induce the cyclodehydrogenation, we annealed the sample stepwise starting at a temperature of 160˚C. The first structural changes appear after annealing to 250˚C. Although the majority of the precursors **1** remain intact at this step, we observed a significant amount of molecules with modified topography. Among them, we could identify intermediates that resemble the unreacted molecule on one side, but are flat on the other side, suggesting a partial cyclodehydrogenation. We assign them to the partially dehydrogenated nanographene **S2** (Figure S11 in the supplementary information). At this stage, however, no precursor **1** reacted completely to form a fully-dehydrogenated nanographene **3**. We annealed therefore the substrate further to 300°C. Subsequently, as shown in Fig. 2a, we observed in the STM images three main different reacted species, whereas two of them (**4**, **5**) are planar and non-symmetric (Fig. 2b,e). To identify the molecular structure of these intermediates, we performed high-resolution STM imaging working with a CO-functionalized tip in the constant-height mode (Fig. 2c,f). This method is known to achieve submolecular resolution, when a molecule is adsorbed flat on the surface.[35]

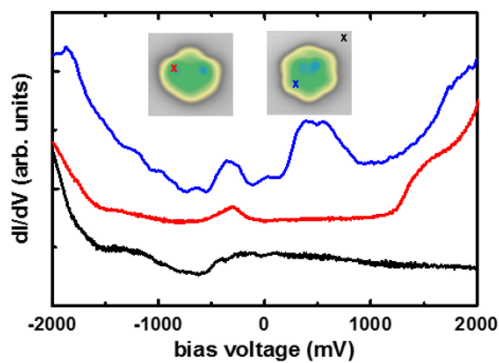

The three different molecular species (**3, 4, 5**) observed after annealing reveal that there is not an exclusive reaction path. The proportion of the respective species was in the present experiment 10% (species **3**), 25% (species **4**) and 65% (species **5**) after annealing to 300 °C (see Fig. S12). The molecule **4** of Fig. 2b appears smaller and, as one can clearly see in the high-resolved STM image of Fig. 2c, the phenyl ring on one side is cleaved, while on the other side two new C-C bonds are formed. This confirms the formation of a nanographene **4** with one five- and one seven-membered ring (Fig. 2d). The molecule **4** is planar because after the detachment of the one phenyl ring, a bending of the benzoisoindole core within the product becomes possible.

A second molecular species **5** appears non-symmetric on the surface (Fig. 2e). From the highly resolved image of Fig. 2f we can conclude that a cyclodehydrogenation reaction has occurred in both upper positions and in one lower position, leading to the formation of two pentagons and one heptagon. This results in some strain in the molecular skeleton, manifested in a discernible curvature of the rings. The least resolved part of the intermediate **5** of Fig. 2f corresponds to a not reacted side of the molecule, which is slightly non-planar because of steric repulsion of the hydrogen atoms. The highest current signal is measured at the position of the native pyrrole ring. This can be understood considering the lone pair of electrons of the nitrogen-atom contributing to the aromaticity of the pyrrole

**Fig. 2** Reaction products after annealing **1** to 300˚C. (a) Overview STM image showing three different molecular species (Imaging parameters: U = -0.6 V, I = 100 pA, 30 nm x 6 nm). (b) STM image of a single molecule **4** with the phenyl ring cleaved on one side (-0.5 V, 100 pA, 2.5 nm x 2.5 nm). (c) Constant-height STM image with CO-functionalized tip of this molecule **4** unambiguously proves the absence of the phenyl ring (bias voltage: 10 mV). Laplace filtering is used to enhance the contrast of the image. (d) Chemical structure of the reaction product shown in (b,c). (e) STM image of a single molecule **5** appearing non-symmetric on-surface (-0.5 V, 100 pA, 2.5 nm x 2.5 nm). (f) Constant-height STM image with CO-functionalized tip of this molecule **5** proves the formation of two pentagons and one heptagon (bias voltage: 10 mV). Laplace filtering is used to enhance the contrast of the image. (g) Chemical structure of the intermediate shown in (e,f).

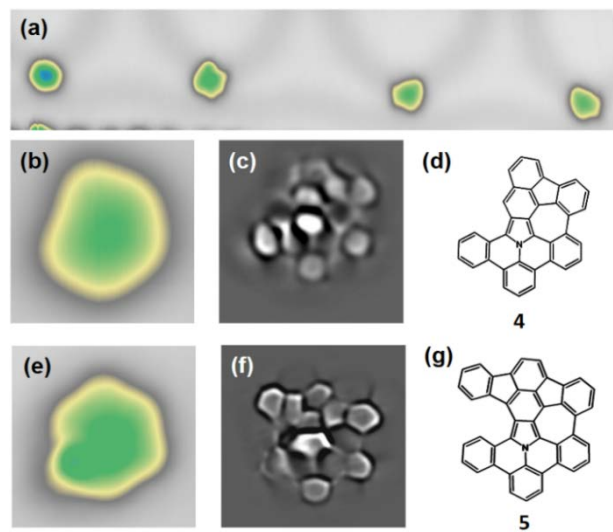

ring. The local electron density is larger on the pyrrole-ring than on the pure carbon rings of the molecular skeleton. This can influence the contrast within the molecule in the STM images.[36] In the STM images after annealing we also observed symmetric molecules with a higher apparent height (left molecule in Fig. 2a). We assign these molecules to the target product **3**. A strain-induced curvature leads to an increased apparent height that is mostly localized in two regions, as can be observed in the STM image in Fig. 3a

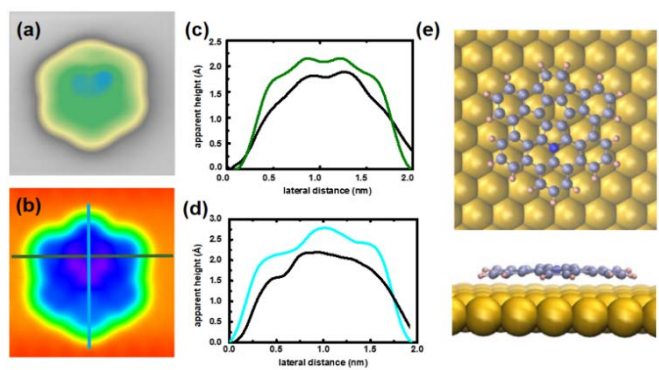

**Fig. 3** The nanographene **3**. (a) STM image of a single fully-reacted molecule (-0.5 V, 100 pA, 2.5 nm x 2.5 nm). (b) Calculated STM image of this species (the same size). (c, d) Linescans along the calculated image in the two directions marked in (b) with the corresponding colours. The black curves correspond to the linescans at the same positions on the experimental STM image (a). (e) Top- and side-view of the molecule adsorbed on the Au(111) surface. The geometry is optimized using DFT. Colour code: gray – carbon, pink – hydrogen, blue – nitrogen.

**Fig. 4** STS spectra of precursor **1** (red curve) and target product **3** (blue curve). dI/dV spectra acquired on the molecules at the positions shown by the crosses in the insets. The colors of the curves correspond to the one of the crosses. The black curve is used as a reference measured on the clean Au(111) surface.

This results in an enhanced interaction between tip and molecule, leading always to a displacement of product 3 on the surface during constant-height measurements (see Fig. S13). The calculated STM image (Fig. 3b) demonstrates the complete reaction of precursor **1** to form a non-planar nanographene **3** with two 5- and 7-membered rings. The comparison of the linescans further confirms the conversion after cyclodehydrogenation. DFT optimization of the target product **3** shows its non-planarity in the upper middle region by about 0.5 Å (Fig. 3e). From these calculations, we can conclude that the curvature is mostly located in the region of the newly formed rings, as the two bright lobes are localized between the pentagon and heptagon, as visible in the linescans of Fig. 3c-d. We investigated the electronic properties of the precursor **1** and of nanographene **3** by comparing the positions of the frontier resonances in STS experiments, thus determining the value of the energy gap (Fig. 4). The observed energies are similar at other positions on the molecule, while only the intensity varies (see Fig. S14). In the precursor **1**, we identified a clear sharp resonance at the bias voltage of -0.32 V and a shoulder with a maximum at 1.45 V for the filled and empty states, respectively.

The energy gap of the precursor **1** can be then estimated to be about 1.77 eV, which is in good agreement with the optical band gap of the *tert*-butyl substituted precursor in solution after considering substrate screening effects.[37] For the nanographene **3**, a new resonance can be clearly identified at about 0.4 V. This represents a significant downshift in comparison to the resonance of the precursor **1**, showing that the formation of the 5-7 membered rings strongly affect the electronic properties of the nanographene molecule **3**, resulting in a new resonance at lower energy.

In conclusion, we have demonstrated the introduction of anomalous 5-7 defects in a nitrogen-containing nanographene by an on-surface cyclodehydrogenation reaction. The reaction pathway goes through a number of intermediate states, with the final product **3** incorporating two azulene motifs and showing a slight curvature on the Au(111) surface. The molecular structure of the reaction intermediates and final product 3 is determined by high-resolution STM measurements with a CO-terminated tip and by theoretical STM image calculations. The electronic properties of nanographene **3** are strongly influenced by the formation of new pentagons and heptagons in the molecular skeleton.

This work has received funding from the German Excellence Initiative *via* the Cluster of Excellence EXC1056 "Center for Advancing Electronics Dresden" (cfaed), the Initiative and Networking Fund of the German Helmholtz Association, Helmholtz International Research School for Nanoelectronic Networks NanoNet (VH-KO-606), the German Research Foundation (*via* SFB 951), and the European Union's Horizon 2020 research and innovation programme under the project MEMO, grant agreement No 766864. D.A.R. thanks the Deutsche Forschungsgemeinschaft (FR2833/50-1, GRK 2247) and the European Graphene Flagship. We further thank the Center for Information Services and High Performance Computing (ZIH) at TU Dresden for computational resources. There are no conflicts to declare.

## Notes and references